\documentclass[aps,english,10pt,tightenlines,a4,twocolumn,superscriptaddress,nofootinbib]{revtex4}  

\usepackage{bm}
\usepackage{bbm}
\usepackage{amsmath}
\usepackage{amsthm}
\usepackage{dsfont}
\usepackage{amssymb}

\usepackage{enumerate}

\usepackage{graphicx}
\usepackage[font={small,it}]{caption}
\usepackage{subcaption}
\usepackage{epstopdf}

\newcommand{\ket}[1]{|#1\rangle}

\newcommand{\ketbra}[2]{|#1\rangle\!\langle#2|}

\newtheorem{theorem}{Theorem}

\newcommand{\id}{\mathbbm{1}}

\DeclareMathOperator{\tr}{tr}

\makeatletter
\newcommand*{\balancecolsandclearpage}{%
  \close@column@grid
  \clearpage
  \twocolumngrid
}
\makeatother

\let\oldmarginpar\marginpar
\renewcommand\marginpar[1]{\-\oldmarginpar[\raggedleft\marginparsize #1]%
{\raggedright\marginparsize #1}}

\newcommand{\CQT}{Centre for Quantum~Technologies, National~University~of~Singapore, 3 Science Drive 2, 117543, Singapore}
\newcommand{\Vie}{Institute for Quantum Optics and Quantum Information,
Austrian Academy of Sciences, Boltzmanngasse 3, A-1090 Vienna, Austria}
\newcommand{\Ont}{Departments of Applied Mathematics and Philosophy,  University of Western Ontario, London, ON N6A 5BY, Canada}
\newcommand{\Peri}{Perimeter Institute for Theoretical Physics, Waterloo, ON N2L 2Y5, Canada}
\newcommand{\ICL}{Blackett Laboratory, Imperial College London, London, SW7 2AZ, United Kingdom}
\newcommand{\Oxf}{Atomic and Laser Physics, Clarendon Laboratory, University of Oxford, Parks Road, Oxford, OX1~3PU, United Kingdom}

\begin{document}

\title{The complex and quaternionic quantum bit \\
 from relativity of simultaneity on an interferometer}

\author{Andrew J.\ P.\ Garner}
\affiliation{\CQT}
\affiliation{\Oxf}
\author{Markus P. M\"uller}
\affiliation{\Vie}
\affiliation{\Peri}
\affiliation{\Ont}
\author{Oscar C.\ O.\ Dahlsten}
\affiliation{\ICL}
\affiliation{\Oxf}
\affiliation{London Institute for Mathematical Sciences, 35a South Street Mayfair, London, W1K 2XF, United Kingdom}

\date{December 11, 2017}

\begin{abstract}
The patterns of fringes produced by an interferometer have long been important testbeds for our best contemporary theories of physics. 
Historically, interference has been used to contrast quantum mechanics to classical physics, but recently experiments have been performed that test quantum theory against even more exotic alternatives.
A physically motivated family of theories are those where the state space of a two-level system is given by a sphere of arbitrary dimension. 
This includes classical bits, and real, complex and quaternionic quantum theory.
In this paper, we consider \emph{relativity of simultaneity} (that observers may disagree about the order of events at different locations) as applied to a two-armed interferometer, 
 and show that this forbids most interference phenomena more complicated than those of complex quantum theory.
If interference must depend on some relational property of the setting (such as path difference), then relativity of simultaneity will limit state spaces to standard complex quantum theory, or a subspace thereof.
If this relational assumption is relaxed, we find one additional theory compatible with relativity of simultaneity: 
 quaternionic quantum theory.
Our results have consequences for current laboratory interference experiments: they have to be designed carefully to avoid rendering beyond-quantum effects invisible by relativity of simultaneity.
\end{abstract}

\maketitle

\section{Introduction}
Questions of locality have historically been used to probe the validity of quantum mechanics.
A famous example is given by the EPR paradox, the subsequent discussions and experimental tests of which contrast the predictions of quantum theory against classical theory~{\cite{EinsteinPR35,Bell64,ClauserHSH69,AspectGR82,WeihsJSWZ98,Hensen15,Giustina15,Shalm15}}.
A relatively new development is to test quantum theory against even more general conceivable non-classical theories (e.g.~\cite{BirkhoffV36,Zeilinger99,Hardy01,Fuchs02,DakicBrukner,Fivel10,ChiribellaDP11,MasanesM11}).

Popescu and Rohrlich have asked whether quantum mechanics has the maximal amount of nonlocality, given its peaceful coexistence with special
relativity~\cite{PopescuR94} (see also~\cite{Tsirelson85,Tsirelson93}), and showed that, surprisingly, the answer is negative: there are conceivable
correlations (now known as \emph{PR-boxes}) that
violate the Bell-CHSH inequality by more than any quantum state, while still not allowing for superluminal information transfer (a property known as \emph{no-signalling}).
This discovery has triggered a whole new area of research, examining the consequences of superstrong nonlocality, and aiming at a simple characterization of quantum correlations~(see e.g.\ \cite{Popescu14} and references within).

In this paper, we apply another relativistic consideration to the fundamental question of whether quantum theory could be modified in some regime (either as a generalisation, or replacement in an as-yet-unencountered limit).
By noting that the number of clicks in a detector should be agreed on by all observers regardless of their frame of reference, we consider the implications of {\em relativity of simultaneity}, where two observers might disagree about the order of events~\cite{Einstein20}.
On interferometers with two spatially separated arms, we show that this principle only allows local transformations
and detector click probabilities that are either identical or very close to the predictions of quantum theory.

We consider the case where the interference pattern is governed by a relational degree of freedom of the device (e.g.\ difference in path lengths),
 such that coherent actions taken on one arm can always be undone by actions on the other.
In this case, we find that consistency with relativity of simultaneity rules out any theory more complicated than standard complex quantum theory.
Thus, the relativistic structure of spacetime itself enforces in this setting that outcome probabilities of measurements are described by the standard rules of quantum theory.
On the other hand, if we relax this strong relational property, then we find that one additional theory is also compatible with relativity of simultaneity: quaternionic quantum theory.

Our results are particularly relevant for a class of experiments that is currently being performed~{\cite{Peres79,KaiserGW84,SinhaCJLW10,SollnerGMPVW11,Weihs13,Procopio17,Kauten17}}, which test the validity of quantum mechanics in specific interferometric setups.
Furthermore, they give interesting insights on the relation between quantum mechanics and spacetime, which is the major object of study in quantum gravity research~\cite{Oriti}.

\balancecolsandclearpage
Borrowing some machinery from the framework of \emph{general probabilistic theories}, we shall begin by introducing a general description for two-level systems, including quantum theory over arbitrary division algebras~\cite{Baez11} (e.g.\ complex numbers, quaternions) and classical bits as special cases.
We then explain how to describe an interferometry setup using this framework.
Next we explicitly introduce relativistic considerations by examining how aspects of this framework should transform under changes of reference frame.
We then consider the restrictions that arise when this is applied to the physical scenario of a two-armed interferometer.
Finally, under different physical background assumptions, we analyse which theories are consistent with relativity in this setup.

\section{General two-level systems}
A classical two-level system (bit) is qualitatively different from a quantum one (qubit).
In the classical case knowing the probability that the system is in one of the two mutually exclusive alternatives is sufficient to totally predict every aspect of its behaviour, whereas in quantum theory this is not the case.
Quantum theory also admits the {\em coherent superposition} of possibilities, allowing for interference effects between actions taken on the two possibilities, as opposed to classical theory that only allows for probabilistic \mbox{mixing}.

The additional information in the qubit beyond the classical two possibilities may be represented by statistics associated with additional {\em complementary measurements}: alternative measurements that could be made on the qubit, whose outcomes cannot be simultaneously predicted (due to the uncertainty principle).
In complex quantum theory, one can find two further complementary measurements in addition to the classical one.
By taking the simultaneously possible expectation values for these three measurements, and representing them as a vector in Euclidean space, one arrives at the three-dimensional Bloch ball.

There has recently been a wave of research results deriving the formalism of quantum theory from simple physical postulates~\cite{Hardy01,DakicBrukner,Fuchs02,Fivel10,ChiribellaDP11,MasanesM11,MullerM13,BarnumMU14}.
In most of these approaches, the first step is to prove that a two-level system is described by a ball state space; simple assumptions on the information-theoretic behaviour of a generalised bit lead to a natural generalization of the three-dimensional Bloch ball: the $d$-dimensional Bloch ball.

Let us now formalise a generalised setup for a system consisting of $d$ such complementary measurements.
The state $\omega$ of such a two-level system is an element of the $d$-dimensional Euclidean unit ball $B^d\equiv\{x\in\mathbb{R}^d\,\,|\,\, |x|\leq 1\}$.
Two-outcome measurements are described by vectors $e\in\mathbb{R}^d$ with $|e|=1$; the probability of the first outcome,
if measured on state $\omega\in B^d$, is $(1+e\cdot \omega)/2$, and that of the second outcome is $(1-e\cdot\omega)/2$.
Transformations which map states to states are given by $d\times d$ orthogonal matrices $R$ acting on $\omega$.
They are \emph{reversible} because by applying $R^{-1}=R^T$, the effect of $R$ can be undone.
In general, one has a compact group $\mathcal{G}\subseteq {\rm O}(d)$ that describes the set of all physically possible reversible transformations on the states.

For $d=1$, the unit ball becomes a line segment, and we recover a classical bit.
$\omega=+1$ and $\omega'=-1$ are the two distinct configurations of a classical spin, and the values in between correspond to probabilistic mixtures. 
The only non-trivial reversible transformation is the bit flip
$R=-\id$, and thus taken together with the trivial transformation $\id$, we see $\mathcal{G}=\mathbb{Z}_2$.

If $d=3$, we recover the two-level systems of standard complex quantum theory: 
Every complex $2\times 2$ density matrix $\rho$ is in one-to-one correspondence with an element $\omega_\rho=(\omega_\rho^1,\omega_\rho^2,\omega_\rho^3)$ of the Bloch ball $B^3$ via $\rho=(\id+\sum_{i=1}^3 \omega_\rho^i \sigma_i)/2$, where $\sigma_i$ are the Pauli matrices.
In this representation, the unitary transformations $\rho\mapsto U\!\rho U^\dagger$ with $U\in {\rm SU}(2)$ are rotations: $\omega_\rho\mapsto R_U\omega_\rho$, where $R_U$ is a suitable element of ${\rm SO}(3)$. 
The probability of outcome $+1$ of a projective measurement with projector $P=|\psi\rangle\langle\psi|$ can be written
$\tr(P\rho)=(1+\omega_P\cdot \omega_\rho)/2$, where $|\omega_P|=1$.
In quantum theory, it is impossible to implement the ``universal NOT'' map $R=-\id$, even though it is a symmetry of the Bloch ball (this corresponds to the transposition of the density matrix, which violates complete positivity).
Thus, the compact group describing the physically possible reversible transformations on a qubit is $\mathcal{G}={\rm SO}(3)$.

The case $d=2$ corresponds to quantum theory of the real numbers; $d=5$ describes a quaternionic quantum bit~\cite{Adler95,Graydon11}, while $d=9$ describes an octonionic two-level system, which can be seen similarly as in the case of complex quantum theory explained above.
The ball state spaces with arbitrary $d\in\mathbb{N}$ have long been known in mathematical physics as examples of state spaces of Jordan algebras~\cite{AlfsenS03}, and they have appeared in various places in quantum information theory~\cite{PawlowskiW12,PaterekDB10,DakicB13}. 
All these state spaces have $N=2$ perfectly distinguishable states and no more~\cite{MullerM13,Hardy01}, with every pair of antipodal points on the sphere (surface of the ball) describing mutually exclusive alternatives.

\section{Generalized interferometry}
Consider the Mach-Zehnder interferometer (MZI) as depicted in Figure 1. A particle (such as a photon) passes through a beam-splitter, travels through the arms, and at the end causes one of the two detectors to click. The click probability depends on the local physical conditions (say, presence of phase plates) in the two branches, with the space-like separated events A and B labelling the moment of passage through the phase plate in question. Each phase plate corresponds to some transformation of the state, $T_A$ and $T_B$ respectively. How can we formally describe this situation without assuming \mbox{quantum} \mbox{mechanics?}

\balancecolsandclearpage

\begin{figure}[!hbt]
\begin{center}
\includegraphics[width=0.4\textwidth]{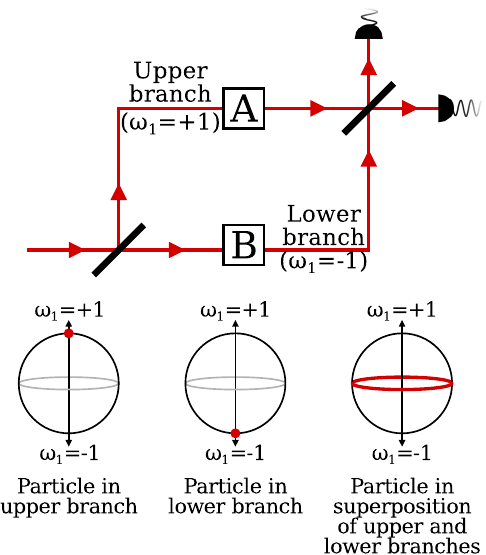}
\caption{
\textbf{A Mach-Zehnder interferometer.}
Upper diagram shows a Mach-Zehnder interferometer, in which a single particle travelling from the bottom left passes through a beam-splitter and enters into a superposition of the two spatially disjoint paths.
The probability that the particle is in either of these paths is quantified by one parameter of the state, $\omega_1$.
At the furthest points in the paths (A and B), two separate agents can choose to insert a phase plate, which might alter the other parameters ($\omega_2, \ldots, \omega_d$).
The paths recombine at the second beam-splitter and the final particle position is measured by a click in a detector.
Lower diagram shows the state of the traversing particle represented on the Bloch ball.
}
\label{fig:BlochMZI}
\end{center}
\end{figure}

First, we can use states $\omega= \left(\omega_1,\ldots,\omega_d\right)$ of a general {\em Bloch ball} state space $B^d$ to describe the location of a single particle in the interferometer, as in Figure~\ref{fig:BlochMZI}.
We can parametrise the state space such that one of the components, say $\omega_1$, determines the probability $p$ to find the particle in the upper branch, as opposed to the lower branch.
According to the general formalism described above, we must have $p=(1+\omega_1)/2$. 
As in quantum theory, there are in general many different states that give the same value $p$.
For example, when $p=1/2$ there can be incoherent mixtures of the two classically possible paths (such as the center of the ball),  coherent superpositions like states on the ball's equator,
or states at any point in between.

Now consider two agents, say Alice and Bob ($A$ and $B$ in Figure~\ref{fig:BlochMZI}), who reside at the spatially separated arms of the interferometer.
If Alice performs an operation on the particle that is reversible and local to her arm (such as inserting a phase plate), then this will be described by some element $T_A$ of the group of physically possible reversible transformations $\mathcal{G}\subseteq {\rm O}(d)$. 
However, since she acts locally, this transformation should not alter the probability $p$ of finding the particle in her branch or the other; otherwise this would allow instantaneous signalling from Alice to Bob.
Thus, $T_A$ must preserve $\omega_1$ for all states.
Hence $T_A\in {\rm O}(d-1)\cap\mathcal{G}$, 
 and the transformation must be an element of the orthogonal group
 associated with the ball of one dimension less, 
 describing all maps that preserve the $\omega_1$-axis.

In order to analyze the setup of Figure~\ref{fig:BlochMZI} in more detail, we need to make some minimal assumptions about how the state space of the particle is related to the physical actions of the beam-splitter and the two agents. For what follows, we will assume the following:
\begin{itemize}
   \item[A1.] By setting up the beam-splitter appropriately, one can prepare a pure state with arbitrary ``upper branch'' probability $p=(1+\omega_1)/2$. 
(This is the state that leaves the beam-splitter.)
   \item[A2.] Every pure state on the Bloch ball $B^d$ with the same upper branch probability $p$ can be prepared by (some combination of) reversible operations applied locally by the agents on the two branches.
(This is the state that will enter the final mirror-detectors system.)
\end{itemize}
In other words, the upper branch probability $p$ is determined at the beam-splitter, and all the other complementary degrees of freedom are set by the local transformations in the arms. 
We assume that this exhausts the full ball state space $B^d$.

Assumptions A1 and A2 preclude the possibility of further, ``hidden'' degrees of freedom of the particle which do not affect any of the measurement outcome probabilities of the setup at all. Without these assumptions, it would be possible, for example, that the ``actual'' state space of the particle (even disregarding additional internal degrees of freedom like spin) is the $(d+1)$-ball $B^{d+1}$, but we happen to see only a $d$-dimensional subspace in our physical setup. Since our goal is to analyze the constraints that arise from the physics of the interferometer alone, we are forced to restrict our consideration to those degrees of freedom that are actually probed by the interferometer. 
Hence, A1 and A2 supply a technical assumption along the lines that the ``effective state space'', the one that is actually probed in this setup, is correctly described by a Euclidean ball $B^d$.

We remark that ball state spaces are a special case in a wider framework known as {\em generalised probabilistic theories} (GPTs)~\cite{Hardy01,Barrett07,MasanesM11}.
In language adopted from this framework,
because the set of $\omega_1$-preserving transformations preserve the statistics associated with the ``which path'' measurement, this set is referred to as the
\emph{phase group} of the ``which path'' measurement~\cite{GarnerDNMV13}.
This definition arises as a natural generalization of the quantum case, where the transformations preserving the statistics of a measurement in a fixed basis $\{\ket{\xi_j}\}_{j=1\ldots N}$ are the unitaries of the form $U(\phi_1, \ldots \phi_N) = \sum_{j=1}^N e^{i\phi_j}\ketbra{\xi_j}{\xi_j}$, and the $\phi_j$ are the \emph{phases} associated with the given transformation.

Even if independent agents act locally on the different branches of the interferometer, their interventions will in general affect some global property of the system (e.g.\ the relative path length), and alter the output statistics of the interferometer (i.e.\ its interference pattern).
However, it remains to be determined if all of the elements in the phase group can be applied by just one agent, or if only a subgroup of these transformations will be available to her.

An operational way to identify the subgroup of the phase transformations associated with an agent acting on a particular branch is to invoke a restriction known as {\em branch~locality}~\cite{DahlstenGV14}, which says that all states with zero probability of being in a particular branch are left invariant under transformations on that branch.
Intuitively, if a particle has no probability whatsoever of travelling down an agent's branch, there should be no way of telling from the output statistics which phase plate (if any) that agent has inserted.
As an example, consider a three-armed interferometer in quantum theory: no transformation applied to the first branch will ever change the relative phase between the second and third branches.

For the spherical state spaces of a two-branch interferometer, branch locality does not introduce any further restrictions: the only state that has zero probability to be found in the upper branch is $\omega=\left(-1,0,\ldots,0\right) \in B^d$, and this state is invariant under all transformations in the phase group ${\rm O}(d-1)\cap \mathcal{G}$.
The same argument applies to the lower branch, hence every single operation in the phase group can be localised to either of the two branches, whilst remaining consistent with branch \mbox{locality}~\cite{DahlstenGV14,Garner14}.

In summary, we conclude that there are two groups%
\footnote{In this paper, we assume that all groups are topologically closed, which implies that they are (as subgroups of the group of orthogonal matrices) Lie groups. This is physically motivated by the fact that we regard a transformation as physically implementable if and only if it can be implemented to arbitrary precision. Due to unavoidable experimental imprecisions, there can be no fundamental difference between perfect and arbitrarily precise implementation.}
$\mathcal{G}_A$ and $\mathcal{G}_B$ of state transformations that describe the result of operations that Alice and Bob respectively can perform locally in their interferometer arms. We have $\mathcal{G}_A,\mathcal{G}_B\subseteq {\rm O}(d-1)\cap\mathcal{G}$, and if we denote the smallest group that contains both $\mathcal{G}_A$ and $\mathcal{G}_B$ by $\mathcal{G}_{AB}$, then our two assumptions tell us that $\mathcal{G}_{AB}$ can map every pure state on the $(d-1)$-ball (representing pure states with equal upper branch probability $p$) to every other one. More technically, this can be formulated as follows:

\begin{eqnarray*}
\mbox{A1,A2} &\Rightarrow& \mathcal{G}_{AB}\mbox{ is transitive on the} \\
 & &  \mbox{surface of the }(d-1)\mbox{-ball.}
\end{eqnarray*}

While we have formalized our description of the two interferometer arms in terms of a single two-level system, one might alternatively try to set up a composite state space $AB$, where $A$ describes the first and $B$ the second arm (such that $AB$ would be some kind of tensor product of both). While the GPT framework allows this in principle, there are several reasons for not doing this here. First, it has been observed before that the usual circuit model, formalizing operations on tensor product spaces directly, is not always the most general or natural model; for example, a physical description of wires in~\cite{Chiribella} allows to implement transformations like the ``switch'' which could otherwise not be implemented directly. This is relevant for our setup, since we are interested in the consequences of our ``wires'' (interferometer arms) being embedded in relativistic spacetime.

\enlargethispage{2\baselineskip}
Second, we are interested in only two exclusive alternatives (upper versus lower arm), and a product state space would necessarily be ``overcomplete'' by containing at least four perfectly distinguishable states. This would have the additional disadvantage that it is currently not clear what the most ``natural'' beyond-quantum four-level state spaces look like, whereas any $d$-dimensional Bloch ball is a natural generalization of the quantum bit. Third, building a composite state space $AB$ would have to rely on auxiliary assumptions like tomographic locality that we consider as unnecessary.

\begin{figure*}[t]
\begin{center}
\includegraphics[width=0.7\textwidth]{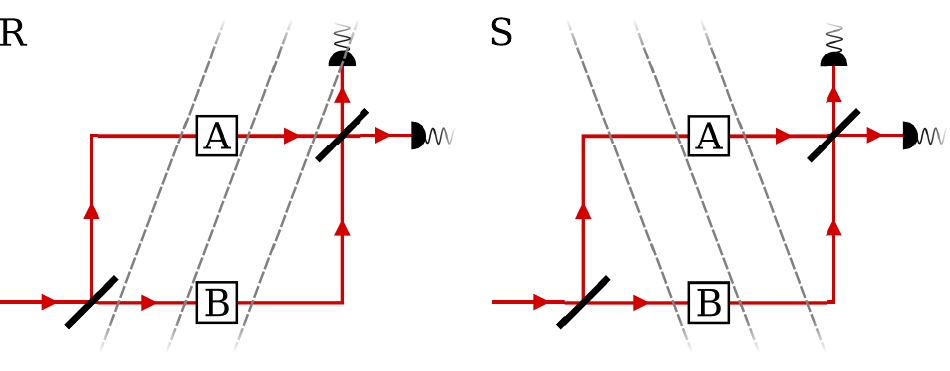}
\caption{
\textbf{A Mach-Zehnder interferometer viewed by two observers.}
Two moving observers Rachael ($R$) and Steven ($S$) witness the operation of an interferometer (described in Figure~\ref{fig:BlochMZI}).
The dashed grey lines show the simultaneous lines of progress for a particle travelling down either branch.
The left diagram shows this as judged by Rachael, the right by Steven.
Rachael therefore witnesses event~A happening before event~B; whereas Steven witnesses the opposite.
If the transformations induced at $A$ and at $B$ did not commute, the observers would disagree about the ultimate detector click statistics.
}
\label{fig:minkowski}
\end{center}
\end{figure*}


\section{Reference frame invariance in the GPT framework}
Local actions of agents in the two interferometer arms are classical space-like separated events. 
Thus, we will analyse the interferometer within special relativity, and consider the effect of a change of reference frame on the general scenario.

The number of clicks in a detector should reflect an objective element of reality. 
One could envision a setup of a photon detector attached to a bomb, such that the world is blown up if and only if the detector clicks; two observers in different frames of reference should not disagree over the fate of the world.
By extension, if different observers do not disagree on whether an event occurred or not (they may still disagree over where, when and in what order relative to other events), they must also agree on the statistics associated with the occurrence of this event.
In our formalism, this means that the probabilities $(1+\omega_P\cdot\omega)/2$ for a positive outcome of measurement $\omega_P$ applied to state $\omega$ should be invariant under a change of reference frame.

Suppose we have two observers, say Rachael and Steven, who both observe the interference experiment, but are moving at relativistic speed relative to each other, as depicted in Figure~\ref{fig:minkowski}. 
Then there will be a Lorentz transformation $\Lambda$ relating Rachael's frame of reference to Steven's. 
A priori, the state $\omega'$ that Steven sees might be different from the state $\omega$ experienced by Rachael, so long as they also describe measurements by different vectors $\omega'_P,\omega_P$ such that the outcome probabilities agree (i.e.\ $\omega_P\cdot\omega
=\omega'_P\cdot\omega'$). 
For this to be possible, there would have to be a representation of the Lorentz group, with reversible transformations $T_\Lambda\in\mathcal{G}$ acting such that $\omega'=T_\Lambda \omega$. 
However, since $\mathcal{G}\subseteq {\rm O}(d)$, this would induce a finite-dimensional unitary representation of the Lorentz group, and it is well-known that the only representation of this kind is the trivial representation~\cite{Wigner39}. 
Moreover, the ``which path'' degree of freedom is not a geometric degree of freedom\footnote{One can imagine that the two material interferometer arms carry classical labels
(like ``$A$'' and ``$B$'', or ``passing Earth'' versus ``passing Mars''). Then different observers will agree on whether a particle is detected in arm $A$ or arm $B$,
and thus on the description of quantum states, regardless of their spacetime
frames of reference.} relative to some Lorentz covariant property, say, a particle's momentum direction (like photon polarization, for example), which is why also Wigner's little group does not apply.
Thus $T_\Lambda=\id$ for all $\Lambda$, and the transformations done by Alice and Bob appear the same in every reference frame.

\section{Consequences of relativity of simultaneity}
We now analyze in detail the idea~\cite{GarnerDNMV13} that relativity of simultaneity might impose further restrictions on the probabilistic behaviour of spatially extended interferometers.
While doing so, we follow the spirit of~\cite{Peres99}, but take Peres' argumentation beyond quantum theory. 
As before, consider the situation in Figure~\ref{fig:minkowski}, with two agents: Alice who acts on the upper branch and Bob who acts on the lower branch.
As the branches are space-like separated, the transformations applied by Alice and Bob should not cause the particle to jump from one branch to another, and so will belong to the phase group $\mathcal{G}_\phi:={\rm O}(d-1)\cap\mathcal{G}$ associated with the ``which path'' measurement. 
More generally, there will be a subgroup of transformations that can be locally applied by Alice, $\mathcal{G}_A \subseteq \mathcal{G}_\phi$, and a subgroup
locally applicable by Bob, $\mathcal{G}_B \subseteq \mathcal{G}_\phi$.

If there is at least one pair of transformations $T_A\in\mathcal{G}_A$ and $T_B\in\mathcal{G}_B$ such that $[T_A,T_B]\neq0$, then the order in which Alice and Bob choose to apply their transformations will have an observable effect on the output statistics of the interferometer at least for some states.
This is particularly problematic for a space-like separation between Alice and Bob:
Consider again the two observers Rachael and Steven, moving at relativistic speed relative to each other, as in Figure~\ref{fig:minkowski}. 
Although Rachael and Steven must agree what effect either action $T_A$ or $T_B$ would have on the interferometer individually (from the Lorentz invariance of transformations),
Steven could in general disagree with Rachael about the order in which the two events occur~\cite{Einstein20}.

In particular, let us say Rachael observes the application of $T_A$ by Alice followed by $T_B$ by Bob; the compound operation is then $T_B T_A$.
When changing into Steven's reference frame, the compound operation should be the same. 
However, Steven may observe instead that $T_B$ happens before $T_A$, describing the compound
operation by $T_A T_B$.
This will lead to contradiction unless $T_A T_B=T_B T_A$ in all cases; that is $[T_A,T_B] = 0$ for all $T_A \in \mathcal{G}_A$ and $T_B \in \mathcal{G}_B$.

In summary, we can formalize an additional assumption that will enter our analysis below:
\begin{itemize}
	\item[REL.] Relativity of simultaneity for the local operations at $A$ and $B$ implies that $[\mathcal{G}_A,\mathcal{G}_B]=0$.
\end{itemize}

Note the subtle way that relativity impacts this interference setup: on the one hand, the ``which path'' information is \emph{not} a geometric degree of freedom that transforms in any non-trivial way under the Lorentz group. 
In other words, the two material interferometer arms break manifest Lorentz symmetry by labelling the branches as ``A'' and ``B'', and all observers see the same transformations $T_A$ and $T_B$.
On the other hand, the local choices or applications of the transformations $T_A$ and $T_B$ are classical space-like separated events, which must not admit a unique time-ordering. Thus, all these transformations $T_A$ and $T_B$ must commute.

\section{Which theories are consistent with relativity?}
We will now analyze which theories are consistent with the assumptions above, in particular with relativity of simultaneity. To this end, we have to make an additional assumption on how Alice's and Bob's groups $\mathcal{G}_A$ and $\mathcal{G}_B$ are related. Due to the symmetry of the setup, it is physically reasonable to expect that $\mathcal{G}_A$ and $\mathcal{G}_B$ are in some sense ``the same''. We will formalize this in two different ways and classify the resulting theories.

\begin{figure*}[!bht]
\begin{center}
\includegraphics[width=0.75\textwidth]{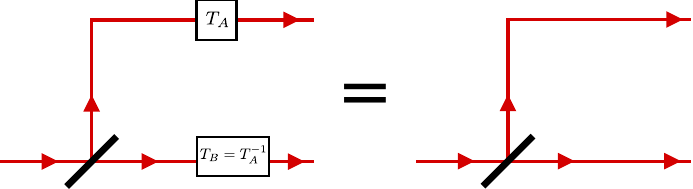}
\caption{
{\textbf{Relational interference.}}
The strong assumption A3* that $\mathcal{G}_A=\mathcal{G}_B$ corresponds to a situation where every transformation on Alice's arm can be ``undone'' by a suitable transformation on Bob's arm (and vice versa). This is the case for the complex quantum bit, but not for the quaternionic quantum bit.
}
\label{fig:inverse}
\end{center}
\end{figure*}

\subsection{Strong assumption: $\mathcal{G}_A=\mathcal{G}_B$}
Let us start with the following assumption (to be relaxed later):
\begin{itemize}
	\item[A3*.] The (groups of) transformations that Alice and Bob can perform locally in their arms are exactly identical: $\mathcal{G}_A=\mathcal{G}_B$.
\end{itemize}
A physical interpretation of this assumption is sketched in Figure~\ref{fig:inverse}: if Alice applies some transformation $T_A\in\mathcal{G}_A$ on her arm, then Bob can always ``undo'' it by performing a suitable transformation $T_B\in\mathcal{G}_B$; since $\mathcal{G}_A=\mathcal{G}_B$ and groups are closed with respect to inversion,  he can simply apply $T_B=T_A^{-1}$.

This behavior can be motivated by the intuition that the local transformations should only change \emph{relational} degrees of freedom; that is, degrees of freedom that correspond to some kind of ``difference'' of properties of the arms that are ``compared'' at the final mirror-detector system. This is the case, for example, in the case of an ordinary optical interferometer: if $T_A$ is implemented by inserting a waveplate into the interferometer arm, then this will change the relative optical path length between the arms. Therefore, it can be undone by simply inserting an identical waveplate in the other arm.

Under this additional assumption, we obtain the following result.
\begin{theorem}
\label{TheComplexQubit}
Under the assumptions A1, A2, A3*, relativity of simultaneity (REL) allows for the following possibilities and not more:
\begin{itemize}
   \item $d=1$ (the classical bit), with $\mathcal{G}_A=\mathcal{G}_B=\{\mathbf{1}\}$ (i.e.\ without any nontrivial local transformations),
	\item $d=2$ (the quantum bit over the real numbers), with $\mathcal{G}_A=\mathcal{G}_B=\mathbb{Z}_2$,
	\item $d=3$ (the standard quantum bit over the complex numbers), with $\mathcal{G}_A=\mathcal{G}_B={\rm SO}(2)={\rm U}(1)$.
\end{itemize}	
Here, $d$ is the dimension of the Bloch ball, and $\mathcal{G}_A$ and $\mathcal{G}_B$ are the local transformations in the interferometer arms.
\end{theorem}
As we will see in the proof, relativity of simultaneity is crucial for this result: without assuming REL, all dimensions $d\in\mathbb{N}$ and a large number of possible groups $\mathcal{G}_A=\mathcal{G}_B$ would be possible. Relativity of simultaneity enforces the which-path degree of freedom to be described by the standard (complex) quantum bit ($d=3$), or a subspace of it (classical bit, or qubit over the reals).
\begin{proof}[Proof of Theorem 1]
	It is easy to check directly that the cases $d=1$ and $d=2$ allow for the groups above and no other ones. The more interesting case is $d\geq 3$ when the surface of $B^{d-1}$, i.e.\ the $(d-2)$-sphere, is a continuous manifold. In this case, not only $\mathcal{G}_{AB}$, but also its connected component at the identity, which we denote $\mathcal{G}_{AB}^0$, must be transitive on the $(d-2)$-sphere~\cite{MasanesMPA2011}. In general, not only the orthogonal groups ${\mathrm O}(d-1)$ and ${\rm SO}(d-1)$ are transitive on the $(d-2)$-sphere $S^{d-2}$, but also subgroups like ${\rm SU}((d-1)/2)$ for odd $d$~\cite{MasanesMPA2011}.
It is possible to exhaustively list the compact connected Lie groups~\cite{MontgomerySamelson,Borel} that act transitively (and effectively\footnote{This means that no two different group elements act in exactly the same way on the sphere. This is a technical assumption that is needed in the mathematical classification results that we are using (otherwise one could always consider the product of a transitive group with another arbitrary group that is supposed to act trivially). In our context, this condition is obviously satisfied, since we \emph{define} the group by its action on the states.}) on $S^{d-2}$, and A1, A2 and A3* imply that $\mathcal{G}_{AB}=\mathcal{G}_A=\mathcal{G}_B$ must be one of them. However, in this infinite list of groups, only one of them is Abelian, as dictated by REL: this is ${\rm U}(1)={\rm SO}(2)$, acting on the surface of $B^{d-1}=B^2$ (the circle).
\end{proof}

In several recent derivations of quantum theory from simple postulates~\cite{MasanesMPA2011,MasanesMPA13}, the condition that ``$\mathcal{G}_{AB}$ is non-trivial and Abelian'' appeared as a crucial mathematical property (though in different context and notation) in the proofs which showed that the Bloch ball must be three-dimensional. 
Here, we obtain an intriguing physical interpretation of this mathematical fact, related to special relativity. 
Furthermore, the derivation above is much easier, and 
 represents one of the simplest arguments
 for why there are three degrees of freedom in a quantum bit\footnote{For another very simple recent derivation of the three-dimensionality of the Bloch ball, see~\cite{Hoehn1,Hoehn2}. A complementary approach to relate the structures of the Bloch ball and of spacetime can be found in~\cite{HoehnMueller}.}.

Clearly, the assumption A3* (i.e.\ that $\mathcal{G}_A=\mathcal{G}_B$), as sketched in Figure~\ref{fig:inverse}, is very strong. 
Let us now therefore relax it.

\subsection{Weaker assumption: $\mathcal{G}_A\simeq \mathcal{G}_B$}
If we look at the symmetry of the interferometric setup, it is  reasonable to expect that the physics is ``the same'' for Alice and Bob: the set of ``phase plates'' (or their beyond-quantum generalizations) available to Alice should be in one-to-one correspondence to the set of phase plates available to Bob. While this still allows that these plates \emph{act differently} on the delocalized particle, it suggests the following assumption (superseding assumption A3*):
\begin{itemize}
	\item[A3.] The transformations that Alice and Bob can perform locally in their arms are isomorphic as topological groups: $\mathcal{G}_A\simeq\mathcal{G}_B$.
\end{itemize}
Similarly as in the previous subsection, we can work out the consequences of A3 and our previous assumptions. We obtain the following generalization of Theorem~\ref{TheComplexQubit}:
\begin{theorem}
Under the assumptions A1, A2, A3, relativity of simultaneity (REL) allows for the following possibilities and not more:
\begin{itemize}
   \item $d=1$ (the classical bit), with $\mathcal{G}_A=\mathcal{G}_B=\{\mathbf{1}\}$ (i.e.\ without any nontrivial local transformations),
	\item $d=2$ (the quantum bit over the real numbers), with $\mathcal{G}_A=\mathcal{G}_B=\mathbb{Z}_2$,
	\item $d=3$ (the standard quantum bit over the complex numbers), with $\mathcal{G}_A=\mathcal{G}_B={\rm SO}(2)={\rm U}(1)$,
	\item $d=5$ (the quaternionic quantum bit), with $\mathcal{G}_{AB}={\rm SO}(4)$, $\mathcal{G}_A$ the left- and $\mathcal{G}_B$ the
right-isoclinic rotations in ${\rm SO}(4)$ (or vice versa) which are both isomorphic to ${\rm SU}(2)$, and $\mathcal{G}_A\cap\mathcal{G}_B=\{+\id, -\id\}$.
\end{itemize}	
As in Theorem~\ref{TheComplexQubit}, $d$ is the dimension of the Bloch ball, $\mathcal{G}_A$ and $\mathcal{G}_B$ are the local transformations in the interferometer arms, and now $\mathcal{G}_{AB}$ is the group generated by all local transformations in $\mathcal{G}_A$ and $\mathcal{G}_B$.
\end{theorem}
That is, a unique additional solution shows up: the quaternionic quantum bit. 
This quaternionic case will necessarily violate the experimental behavior sketched in Figure~\ref{fig:inverse}: 
Except for the reflection map $-\id$ (and the identity map $\id$ itself), no other of Alice's local operations can be undone by Bob.
However, the ability to undo just these two operations is sufficiently permissive to allow the $d=5$ interferometer to implement the Deutsch-Jozsa algorithm~\cite{Garner16}, suggesting that this additional case is computationally interesting.

\begin{proof}[Proof of Theorem 2]
If $\mathcal{G}_A=\mathcal{G}_B$ then we are back in the case that is treated in Theorem~\ref{TheComplexQubit}, leading to the first three cases $d=1,2,3$ listed above (and no other ones). Let us therefore assume that $\mathcal{G}_A\neq\mathcal{G}_B$, which implies in particular that $\mathcal{G}_B$ contains more than just the identity element. We may also assume that $d\geq 3$, since we have already enumerated all the cases with $d=1,2$. It is easy to see that the commutant
\[
   \mathcal{G}'_A:=\{G\in\mathcal{G}_{AB}\,\,|\,\, GX=XG\mbox{ for all }X\in\mathcal{G}_A\}
\]
is a normal subgroup of $\mathcal{G}_{AB}$.
Consider first the case $\mathcal{G}'_A=\mathcal{G}_{AB}$.
Since $\mathcal{G}_A\subseteq\mathcal{G}_{AB}$, this implies that $\mathcal{G}_A$ is Abelian, and then A3 implies that $\mathcal{G}_B$ is Abelian too. Due to REL, it follows that arbitrary products of elements of $\mathcal{G}_A\cup\mathcal{G}_B$ can be ordered in arbitrary ways, which implies that $\mathcal{G}_{AB}$ must be Abelian too. But A1 and A2 imply that $\mathcal{G}_{AB}$ is transitive on the $(d-2)$-sphere, and then we are back in the case discussed in the proof of Theorem~\ref{TheComplexQubit}: only the case of the standard complex quantum bit, $d=3$, is possible.

Now consider the second case $\mathcal{G}'_A\subsetneq \mathcal{G}_{AB}$, and let $\mathcal{G}_{AB}^0$ be its connected component at the identity, which must then also be transitive on the $(d-2)$-sphere due to A1 and A2. 
We may also assume that $\mathcal{G}_{AB}^0$ is non-Abelian, since otherwise we fall back into the previous case. 
REL implies that $\mathcal{G}_B\subseteq \mathcal{G}'_A$, thus $\mathcal{G}'_A$ is non-trivial. Suppose that $\mathcal{G}_B$ was a discrete group, then so would be $\mathcal{G}_A$, and since $\mathcal{G}_{AB}\subseteq \{T_A T_B\,\,|\,\,T_A\in\mathcal{G}_A,T_B\in\mathcal{G}_B\}$ due to REL, this would imply that $\mathcal{G}_{AB}$ is discrete too, contradicting its transitivity on the $(d-2)$-sphere (and hence contradicting A1 and A2). Therefore $\mathcal{G}_B$ is not discrete, hence $\mathcal{G}'_A$ has a non-trivial connected component at the identity, $\mathcal{G}'_{A,0}$.
It is easy to see that $\mathcal{G}'_{A,0}$ inherits normality from $\mathcal{G}'_A$. That is, $\mathcal{G}'_{A,0}$ is a non-trivial connected proper \mbox{normal} subgroup of $\mathcal{G}_{AB}$, and thus of $\mathcal{G}_{AB}^0$. In other words, $\mathcal{G}_{AB}^0$ is not a simple Lie group, and it is also non-Abelian.

Looking again at the list of compact connected Lie groups that act transitively and effectively on the spheres, this leaves only the following possibilities for $\mathcal{G}_{AB}^0$: ${\rm SO}(4)$ for $d=5$, and essentially%
\footnote{The term ``essentially'' refers to the fact that we have to divide this group by a finite subgroup to obtain an effective group action, see~\cite{MontgomerySamelson}.} 
${\rm Sp}((d-1)/4)\times {\rm U}(1)$ for $d-1=8,12,16\ldots$ as well as essentially ${\rm Sp}((d-1)/4)\times {\rm SU}(2)$ for $d-1=4,8,12,\ldots$. 
Since the Lie algebras of ${\rm SO}(4)$ and ${\rm Sp}((d-1)/2)\times {\rm SU}(2)$ are semisimple, the decomposition of these Lie algebras into ideals is unique, and thus the sets of normal connected Lie subgroups of these groups can be read off directly (in particular, the symplectic groups are simple~\cite{MontgomerySamelson}). If $\mathcal{G}_{AB}^0={\rm SO}(4)$ then $\mathcal{G}'_{A,0}$ must be either the left- or the right-isoclinic rotations in ${\rm SO}(4)$ since these are the only non-trivial connected normal subgroups. Suppose $\mathcal{G}'_{A,0}={\rm SO}(4)_R$, the right-isoclinic rotations (otherwise relabel $A\leftrightarrow B$). 
Then $\mathcal{G}'_A\supseteq {\rm SO}(4)_R$, and so every $X\in\mathcal{G}_A$ must commute with every $G\in{\rm SO}(4)_R$. It is easy to see that no reflection $X\in{\rm O}(4)$ with $\det X=-1$ can have this property; among the rotations, only the left-isoclinic rotations satisfy this. Thus $\mathcal{G}_A\subseteq {\rm SO}(4)_L$. 
Since $\mathcal{G}_A\simeq \mathcal{G}_B$ this implies that $\mathcal{G}_B$ does not contain any reflections either, and so $\mathcal{G}_{AB}=\mathcal{G}_{AB}^0={\rm SO}(4)$. Furthermore, this implies that $\mathcal{G}_B\subseteq\mathcal{G}'_A=\mathcal{G}'_{0,A}={\rm SO}(4)_R$. 
However, if $\mathcal{G}_A$ (or $\mathcal{G}_B$) were proper Lie subgroups of ${\rm SO}(4)_R$ (resp.\ ${\rm SO}(4)_L$) then they would be too small to generate $\mathcal{G}_{AB}$.
 We have thus recovered the quaternionic quantum bit, i.e.\ the $d=5$ case above.

If $\mathcal{G}_{AB}^0$ is essentially ${\rm Sp}((d-1)/4)\times {\rm SU}(2)$ then one of $\mathcal{G}'_{A,0}$ or $\mathcal{G}'_{B,0}$ must correspond to ${\rm SU}(2)$, hence both $\mathcal{G}_A$ and $\mathcal{G}_B$ must be a subgroup of ${\rm SU}(2)$. Since $\mathcal{G}_A$ and $\mathcal{G}_B$ generate $\mathcal{G}_{AB}$, this is only possible if $d=5$, in which case we recover the quaternionic case again, since ${\rm Sp}(1)\simeq {\rm SU}(2)$ and ${\rm SU}(2)\times {\rm SU}(2)/\mathbb{Z}_2\simeq{\rm SO}(4)$.

Finally, for the case that $\mathcal{G}_{AB}^0$ is essentially ${\rm Sp}((d-1)/4)\times {\rm U}(1)$, it is not difficult to see that all ideals of the corresponding Lie algebra are either trivial or one-dimensional, so that no two non-trivial normal Lie subgroups are sufficient to generate all of $\mathcal{G}_{AB}$.
\end{proof}

\section{Implications for interference experiments}
Recent experimental activity has aimed at testing quantum mechanics against more general theories in interference experiments~\cite{SinhaCJLW10,SollnerGMPVW11}. 
One proposal~\cite{Weihs13} is due to Peres~\cite{Peres79}, in which an interferometric setup discriminates between ordinary quantum theory and its quaternionic counterpart, or more generally, whether a quantum two-level system has more than three complementary measurements. Our results provide further motivation to test for quaternionic quantum mechanics ($d=5$).
Our analysis also provides some insights into the operational behavior of such conceivable quaternionic effects: they would have to test fundamentally ``non-relational'' but still non-localized degrees of freedom of a particle, as sketched in Figure~\ref{fig:inverse}. 
Such quantities cannot correspond to differences of scalar quantities like path lengths. 
Conceptually, this resembles the insights obtained in a different formalism~\cite{BGW}, showing that composition of quaternionic quantum state spaces must violate tomographic locality to have the ``nice'' properties that one would physically expect.

There has been at least one recent experimental implementation which is in principle able to detect effects of Bloch ball dimension $d\geq 6$~\cite{Kauten17,TalkBrukner}. Our results suggest that such conceivable exotic effects are even less likely to be detected: the principle of relativity of simultaneity is sufficient to rule them out. In this paper, we have only started to explore the constraints that relativistic causality imposes on probabilistic theories, and it seems likely that more constraints will be found when we go to interferometers with more than two arms, or look at further consequences of Minkowski spacetime beyond relativity of simultaneity. This may suggest to perform interference experiments in such a way that they are not subject to these constraints. For example, one might implement the mutually exclusive alternatives (which are supposed to interfere) as energy levels or orthogonal spin directions, instead of spacelike separated slits.

Suppose we keep the assumption $\mathcal{G}_A=\mathcal{G}_B$ which expresses the relational character of the phase transformations (which is violated by the quaternionic qubit), but drop assumptions A1 and A2 so that we are not restricted to Bloch ball state spaces. Relativity of simultaneity still enforces that this group is Abelian. Then elementary group theory~\cite{Sepanski07} shows that $\mathcal{G}_A=\mathcal{G}_B$ must be of the form
${\rm SO}(2)\oplus {\rm SO}(2)\oplus\ldots\oplus {\rm SO}(2)$, with some of the $n$ addends possibly replaced by a finite subgroup. 
For example, quaternionic quantum bits ($d=5$) have an Abelian subgroup of phase dynamics isomorphic to the torus ${\rm SO}(2)\oplus{\rm SO}(2)$. 
Local transformations in the interferometer arms will then be described by $n$ independent complex phases,
 such that the interferometer is operationally similar to $n$ independent complex quantum interferometers (with the added restriction that only one of the quantum interferometers can be measured per run of the experiment){~\cite{Garner14}}.
This suggests that the experimental detection of possible deviations will be very difficult.

\section{Summary and Outlook}
In this paper, we have considered how relativity constrains observable interference phenomena, without assuming that the probabilities of detector clicks are necessarily described by quantum mechanics.
We have shown that relativity of simultaneity on two-armed interferometers enforces a behaviour that is very close to that of standard complex quantum theory.
If the two arms correspond to mutually exclusive alternatives of a $d$-dimensional ball state space, then there are essentially two possibilities: either the the state space is given by the standard complex quantum bit ($d=3$) and its subspaces, or by the quaternionic qubit ($d=5$). These two cases behave very differently: the nonlocal degrees of freedom of the quaternionic qubit cannot be interpreted as purely relational, in contrast to the standard qubit.

Two questions arise: 
(i) given the findings above, what consequences can we draw for implementations of experiments like those that test for higher-order interference~\cite{SinhaCJLW10, SollnerGMPVW11, Weihs13, Procopio17, Kauten17} in Sorkin's sense~\cite{Sorkin}? 
While we have been able to draw some conclusions for experiments like Peres'~\cite{Peres79}, addressing these other types of experiments demands to extend our analysis beyond two interferometer arms.
(ii) How far can one push this type of reasoning: is quantum theory the only theory consistent with relativistic causality?
Equivalently: is the quantum path integral rule for summing up complex phases a direct consequence of the structure of spacetime? 
Our results suggest this may well be the case.

\section*{ACKNOWLEDGEMENTS}
We are grateful for illuminating discussions and correspondence~with {\v{C}aslav~Brukner}, {Borivoje~Daki\'c}, {Nana~Liu}, \mbox{Felix~Pollock}, and \mbox{Benjamin~Yadin}. 
Furthermore, we acknowledge financial support from the EPSRC (UK), the John Templeton Foundation, and the Foundational Questions Institute.


\end{document}